\documentclass[sigconf]{acmart}
\usepackage{graphicx}
\usepackage[ruled]{algorithm2e}
\usepackage[utf8]{inputenc}
\usepackage{subfigure}
\usepackage{multirow}
\usepackage{tabulary}
\usepackage{amsthm}
\usepackage{algorithmic}
\usepackage{enumitem}
\usepackage{epstopdf}

\AtBeginDocument{%
  \providecommand\BibTeX{{%
    \normalfont B\kern-0.5em{\scshape i\kern-0.25em b}\kern-0.8em\TeX}}}

\begin{document}
\title{PipAttack: Poisoning Federated Recommender Systems for Manipulating Item Promotion}
\author{Shijie Zhang}
\affiliation{%
  \institution{The University of Queensland}
  \country{}}
\email{shijie.zhang@uq.edu.au}

\author{Hongzhi Yin}
\authornote{Corresponding author.}
\affiliation{%
  \institution{The University of Queensland}
  \country{}}
\email{h.yin1@uq.edu.au}
\author{Tong Chen}
\affiliation{%
  \institution{The University of Queensland}
  \country{}}
\email{tong.chen@uq.edu.au}
\author{Zi Huang}
\affiliation{%
  \institution{The University of Queensland}
  \country{}}
\email{huang@itee@uq.edu.au}
\author{Quoc Viet Hung Nguyen}
\affiliation{%
  \institution{Griffith University}
  \country{}}
\email{quocviethung.nguyen@griffith.edu.au}
\author{Lizhen Cui}
\affiliation{%
  \institution{Shandong University}
  \country{}}
\email{clz@sdu.edu.cn}

\begin{abstract}
Due to the growing privacy concerns, decentralization emerges rapidly in personalized services, especially recommendation. Also, recent studies have shown that centralized models are vulnerable to poisoning attacks, compromising their integrity. In the context of recommender systems, a typical goal of such poisoning attacks is to promote the adversary's target items by interfering with the training dataset and/or process. Hence, a common practice is to subsume recommender systems under the decentralized federated learning paradigm, which enables all user devices to collaboratively learn a global recommender while retaining all the sensitive data locally. Without exposing the full knowledge of the recommender and entire dataset to end-users, such federated recommendation is widely regarded `safe' towards poisoning attacks. In this paper, we present a systematic approach to backdooring federated recommender systems for targeted item promotion. The core tactic is to take advantage of the inherent popularity bias that commonly exists in data-driven recommenders. As popular items are more likely to appear in the recommendation list, our innovatively designed attack model enables the target item to have the characteristics of popular items in the embedding space. Then, by uploading carefully crafted gradients via a small number of malicious users during the model update, we can effectively increase the exposure rate of a target (unpopular) item in the resulted federated recommender.  Evaluations on two real-world datasets show that 1) our attack model significantly boosts the exposure rate of the target item in a stealthy way, without harming the accuracy of the poisoned recommender; and 2) existing defenses are not effective enough, highlighting the need for new defenses against our local model poisoning attacks to federated recommender systems.
\end{abstract}


\maketitle
\vspace{-0.5em}
\section{Introduction}\label{sec:intro}
\begin{figure*}[t!]
\centering
\begin{tabular}{cccc}
\hspace{0.5em}\includegraphics[width=1.3in]{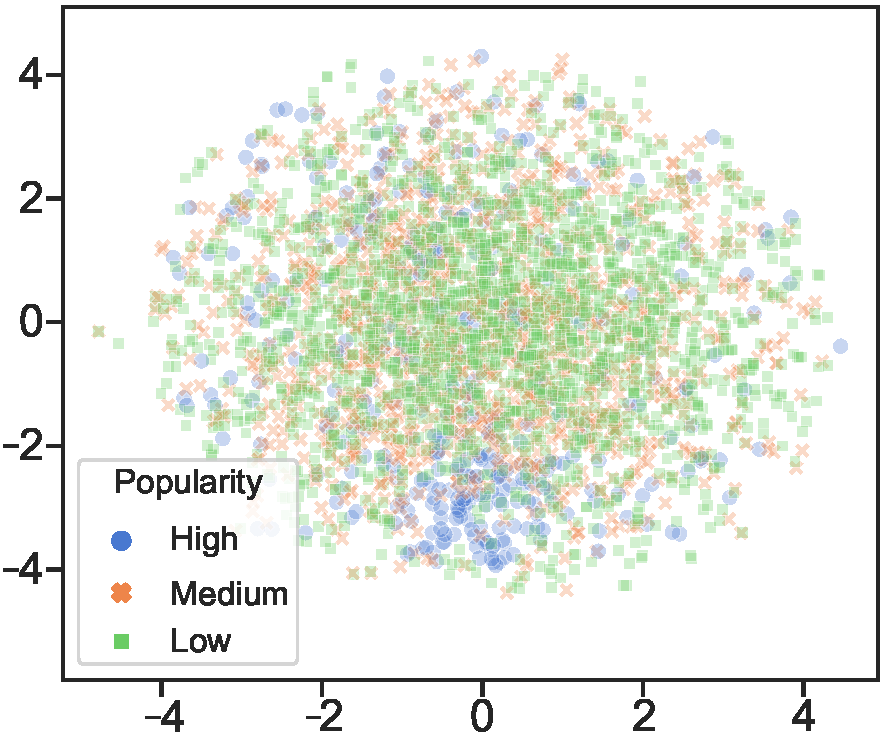}
&\hspace{2.5em}\includegraphics[width=1.3in]{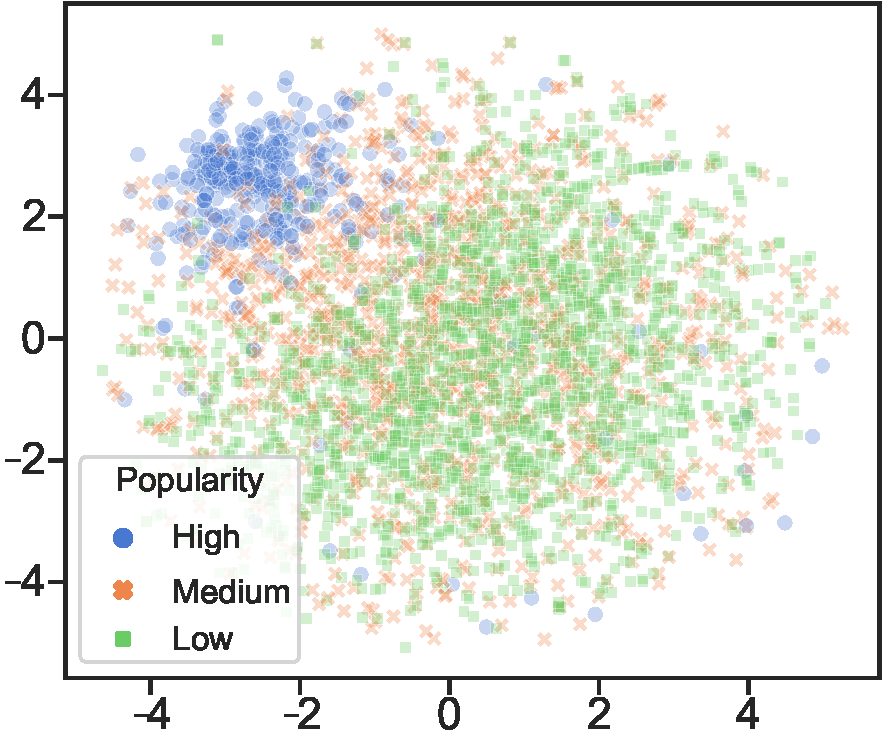}
&\hspace{2.5em}\includegraphics[width=1.27in]{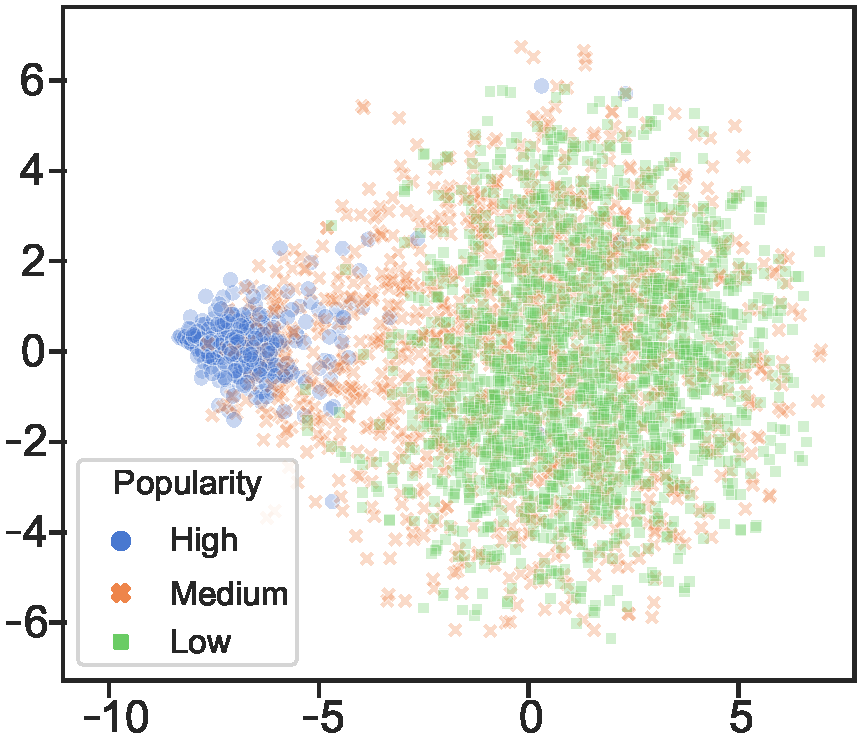}
&\hspace{-0.5em}\includegraphics[width=1.55in]{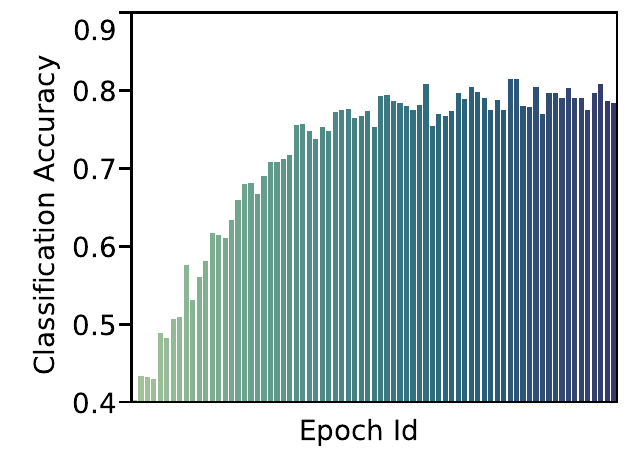}\\
\small{(a) 5th Epoch} &\hspace{2em}\small{(b) 20th Epoch} &\hspace{2em}\small{(c) 40th Epoch} &\small{(d) Popularity Classification Accuracy (F1)}\\
\end{tabular}
\vspace{-1.2em}
\caption{Visualization on the popularity bias of a CF-based federated recommender. (a), (b) and (c) are the t-SNE projection of item embeddings at different training stages, where more details on popularity labels can be found in Section~\ref{lb:data}. (d) shows the convergence curve of a popularity classifier using $F1$ score.}
\label{fig:stas}
\vspace{-1.2em}
\end{figure*}
The demand for recommender systems has increased more than ever before. Over the years, various recommendation algorithms have been proposed and proven successful in various applications. Collaborative filtering (CF), which infers users' potential interests from users' historical behavior data, lies at the core of modern recommender systems. Recently, enhancing CF with deep neural networks has been a widely adopted means for modelling the complex user-item interactions~\cite{he2017neural} and demonstrated state-of-the-art performance in a wide range of recommendation tasks.

The conventional recommender systems centrally store users' personal data to facilitate the centralized model training, which are, however, increasing the privacy risks~\cite{10.1145/3159652.3159688}. Apart from privacy issues, another major risk emerges w.r.t. the correctness of the learned recommender in the presence of malicious users, where the trained model can be induced to deliver altered results as the adversary desires, e.g., promoting a target product\footnote{Product demotion works analogously, hence we mainly focus on promoting an unpopular product in this paper.} such that it gets more exposures in the item lists recommended to users. This is termed \textbf{\textit{poisoning attack}} \cite{lin2020attacking,fang2020influence}, which is usually driven by financial incentives but incurs strong unfairness in a trained recommender. In light of both privacy and security issues, there has been a recent surge in decentralizing recommender systems, where federated learning \cite{wu2021fedgnn,muhammad2020fedfast, wang2021fast} appears to be one of the most representative solutions. Specifically, a federated recommender allows users' personal devices to locally host their data for training, and the global recommendation model is trained in a multi-round fashion by submitting a subset of locally updated on-device models to the central server and performing aggregation (e.g., model averaging).

Subsuming a recommender under the federated learning paradigm naturally protects user privacy as the data is no longer uploaded to a central server or cloud. Moreover, it also prevents a recommender from being poisoned. The rationale is, in recommendation, the predominant poisoning method for manipulating item promotion is data poisoning, where the adversary manipulates malicious users to generate fake yet plausible user-item interactions (e.g., ratings), making the trained recommender biased towards the target item~\cite{zhang2020practical,lin2020attacking,fang2018poisoning,fang2020influence}. However, these aforementioned data poisoning attacks operate under the assumption that the adversary has prior knowledge about the entire training dataset. Apparently, such assumption is voided in federated recommendation as the user data is distributively held, thus restricting the adversary's access to the entire dataset. Though the adversary can still conduct data poisoning by increasing the amount of fake interactions and/or malicious user accounts, it inevitably makes the attack more identifiable from either the abnormal user behavioral footprints \cite{zhang2020gcn,kumar2018rev2} or heavily impaired recommendation accuracy \cite{lin2020attacking}.

Thus, we aim to answer this challenging question: \textit{can we backdoor federated recommender systems via poisoning attacks?} In this paper, we provide a positive answer to this question with a novel attack model that manipulates the target item's exposure rate in this non-trivial decentralized setting. Meanwhile, our study may also shed some light on the security risks of federated recommenders. Unlike data poisoning, our attack approach is built upon the model poisoning scheme, which compromises the integrity of the model learning process by manipulating the gradients~\cite{bhagoji2019analyzing,baruch2019little} of local models submitted by several malicious users. Given a federated recommender, we assume an adversary controls several malicious users/devices, each of which has the authority to alter the local model gradients used for updating the global model. Since the uploaded parameters directly affect the global model, it is more cost-effective for an adversary to poison the federated recommender from the model level, where a small group of malicious users will suffice.

Despite the efficacy of model poisoning in many federated learning applications, the majority of these attacks are only focused on perturbing the results in classification tasks (e.g., image classification and word prediction)~\cite{bagdasaryan2020backdoor,fang2020local}. However, designed for personalized ranking tasks, federated recommenders are optimized towards completely different learning objectives, leaving poisoning attacks for item promotion largely unexplored. In the meantime, the federated environment significantly limits our adversary's prior knowledge to only partial local resources (i.e., malicious users' data and models) and the global model. Moreover, the poisoned global model should maintain high recommendation accuracy, so as to promote the target item stealthily while providing high-quality recommendations to benign users.

To address these challenges, we take advantage of a common characteristics of recommendation models, namely the popularity bias.
As pointed out by previous studies \cite{zhu2020measuring,zhu2021popularity,zhang2021causal,abdollahpouri2017controlling}, data-driven recommenders, especially CF-based methods are prone to amplify the popularity by over-recommending popular items that are frequently visited. Intuitively, because such inherent bias still exists in federated recommenders, if we can disguise our target item as the popular items by uploading carefully crafted local gradients via malicious users, we can effectively trick the federated recommender to become biased towards our target item, thus boosting its exposure. Specifically, we leverage the learnable item embeddings as an interface to facilitate model poisoning. To provide a proof-of-concept, in Figure~\ref{fig:stas}, we use t-SNE to visualize the item embeddings learned by a generic federated recommender (see Section \ref{sec:base_rec} for model configuration). As the model progresses from its initial state towards final convergence (Figure~\ref{fig:stas}(a)-(c)), items from three different popularity groups gradually forms distinct clusters. To further reflect the strong popularity bias, we train a simplistic popularity classifier (see Section \ref{sec:pop_obf}) that predicts an item's popularity group given its learned embeddings, and show its testing performance in Figure~\ref{fig:stas}(d). In short, the high prediction accuracy ($F1 >0.8$) again verifies that the item embeddings in a well-trained federated recommender are highly discriminative w.r.t. their popularity.

To this end, we propose \underline{\textbf{p}}oisoning \underline{\textbf{attack}} for \underline{\textbf{i}}tem \underline{\textbf{p}}romotion (PipAttack), a novel poisoning attack model targeted on recommender systems in the decentralized, federated setting. Apart from optimizing PipAttack under the \textit{explicit promotion constraint} that straightforwardly pushes the recommender to raise the ranking score of the target item, we innovatively design two learning objectives, namely the \textit{popularity obfuscation constraint} and \textit{distance constraint} so as to achieve our attack goal via fewer model updates without imposing dramatic accuracy drops. On one hand, popularity obfuscation confuses the federated recommender by aligning the target item with popular ones in the embedding space, thus greatly benefiting the target item's exposure rate via the popularity bias. On the other hand, to avoid harming the usability of the federated recommender and being detected, our attack model bears a distance constraint to prevent the manipulated gradients uploaded by malicious users from deviating too far from the original ones. We summarize our main contributions as follows:

\begin{itemize}
    \item To the best of our knowledge, we present the first systematic approach to poisoning federated recommender systems, where an adversary only has limited prior knowledge compared with existing centralized scenarios. Our study reveals the existence of recommenders' security backdoors even in a decentralized environment.
    \item We propose PipAttack, a novel attack model that induces the federated recommender to promote the target item by uploading carefully crafted local gradients through several malicious users. PipAttack assumes no access to the full training dataset and other benign users' local information, and innovatively takes advantage of the inherent popularity bias to effectively boost the exposure rate of the target item.
    \item Experiments on two real-world datasets demonstrate the advantageous performance of PipAttack even when defensive strategies are in place. Furthermore, compared with all baselines, PipAttack is more cost-effective as it can reach the attack goal with fewer model updates and malicious users.
\end{itemize}
\vspace{-0.5em}
\section{Preliminaries}
In this section, we first revisit the fundamental settings of federated recommendation and then formally define our research problem.
\vspace{-0.5em}
\subsection{Federated Recommender Systems}
Let $\mathcal{V}$ and $\mathcal{U}$ denote the sets of $N$ items and $M$ users/devices, respectively. Each user $u_i\in \mathcal{U}$ owns a local training dataset $\mathbf{\mathcal{D}}_{i}$ consisting of implicit feedback tuples $(u_i, v_j, r_{ij})$, where $r_{ij} = 1$ if $u_i$ has visited item $v_j \in \mathcal{V}$ (i.e., a positive instance), and $r_{ij} = 0$ if there is no interaction between them (i.e., a negative instance). Note that the negative instances are downsampled using a positive-to-negative ratio of $1:q$ for each $u_i$ due to the large amount of unobserved user-item interactions. Then, for every user, a federated recommender (FedRec) is trained to estimate the feedback $\hat{r}_{ij} \in [0,1]$ between $u_i$ and all items, where $\hat{r}_{ij}$ is also interpreted as the ranking/similarity score for a user-item pair. With the ranking score computed, FedRec then recommends an item list for each user $u_i$ by selecting $K$ top-ranked items w.r.t. $\hat{r}_{ij}$. It can be represented as:
\begin{equation}
    FedRec(u_i|\Theta) = \{v_j | \hat{r}_{ij}\,\, \textnormal{is top-}K \,\, \textnormal{in} \,\, \{\hat{r}_{ij'}\}_{j'\in \bar{\mathcal{I}}(i)} \},
\end{equation}
where $\bar{\mathcal{I}}(i)$ denotes unrated items of $u_i$ and $\Theta$ denotes all the trainable parameters in FedRec. For notation simplicity, we directly use $\Theta$ to represent the recommendation model.

\textbf{Federated Learning Protocol.} In FedRec, a central server coordinates individual user devices, of which each keeps $\mathbf{\mathcal{D}}_{i}$ and a copy of the recommendation model locally. To train FedRec, the local model on the $i$-th user device is optimized locally by minimizing the following loss function:
\begin{equation}
\resizebox{.9\linewidth}{!}{$
    \mathcal{L}^{rec}_i = - \sum_{(u_i,v_j, r_{ij})\in \mathcal{D}_i} r_{ij}\log \hat{r}_{ij} + (1-r_{ij})\log(1-\hat{r}_{ij}),
    $}
\label{eq:rating}
\end{equation}
where we treat the estimated $\hat{r}_{ij} \in [0,1]$ as the probability of observing the interaction between $u_i$ and $v_j$. Then, the above cross-entropy loss quantifies the difference between $\hat{r}_{ij}$ and the binary ground truth ${r}_{ij}$. It is worth mentioning that, other popular distance-based loss functions (e.g., hinge loss and Bayesian personalized ranking loss \cite{rendle2012bpr}) are also applicable and behave similarly in FedRec. With the user-specific loss $\mathcal{L}_i^{rec}$ computed, we can derive the gradients of $u_i$'s local model $\Theta_i$, denoted by $\nabla \Theta_i$. At iteration $t$, a subset of users $\mathcal{U}_t$ are randomly drawn. Each $u_i\in\mathcal{U}_t$ then downloads the latest global model $\Theta_t$ and updates its local gradients w.r.t. $\mathcal{D}_i$, denoted by $\nabla \Theta^t_{i}$. After the central server receives all local gradients submitted by $|\mathcal{U}_t|$ users, it aggregates the collected gradients to facilitate global model update. Specifically, FedRec follows the commonly used gradient averaging \cite{mcmahan2017communication} to obtain the updated model $\Theta_{t+1}$ with learning rate $\eta$:
\begin{equation}
\resizebox{0.5\linewidth}{!}{$
	\Theta_{t+1} = \Theta_{t} - \eta \frac{1}{|\mathcal{U}^t|} \sum_{u_i \in \mathcal{U}_t}\nabla \Theta_i^{t}.
	$}
\end{equation}
The training proceeds iteratively until convergence. Unlike centralized recommenders, FedRec collects only each user's local model gradients instead of her/his own data $\mathcal{D}_i$, making existing poisoning attacks \cite{lin2020attacking, christakopoulou2019adversarial,zhang2020practical} fail due to their ill-posed assumptions on the access to the entire dataset. Furthermore, different local gradients are not shared across users, which further reduces the amount of prior knowledge that a malicious party can acquire.
\vspace{-0.5em}
\subsection{Poisoning Attacks}
Following \cite{fang2020local}, we assume an adversary can compromise a small proportion of users in FedRec, which we term malicious users. The attack is scheduled to start at epoch $t$, and all malicious users participate in the training of FedRec normally before that. From epoch $t$, the malicious users start poisoning the global model by replacing the real local gradients $\nabla \Theta_i^t$ with purposefully crafted gradients $\widetilde {\nabla \Theta_i^t}$, so as to gradually guide the global model to recommend the target item more frequently. The task is formally defined as follows:

\textit{Problem 1.} \textbf{Poisoning FedRec for Item Promotion.} Given a federated recommender parameterized by $\Theta$, our adversary aims to promote a target item $v^*_j \in \mathcal{V}$ by altering the local gradients submitted by every compromised malicious user device $u_i^* \in \mathcal{U}^*$ at each update iteration $t$, i.e., $f:\nabla \Theta_i^t \mapsto \widetilde{\nabla \Theta_i^t}$. Note that each malicious user will produce its unique deceptive gradients $\widetilde{\nabla \Theta_i^t}$. Hence, the deceptive gradients $\{\widetilde{\nabla \Theta_i^t}\}_{\forall u_i \in \mathcal{U}^*}$ for all malicious users are parameters to be learned during the attack process, such that for each benign user $u_i \in \mathcal{U}\backslash \mathcal{U}^*$, the probability that $v^*_j$ appears in $FedRec(u_i|\Theta)$ is maximized.
\vspace{-0.5em}
\section{Poisoning FedRec for Item Promotion}
\begin{figure}
\centering
\includegraphics[scale=0.365]{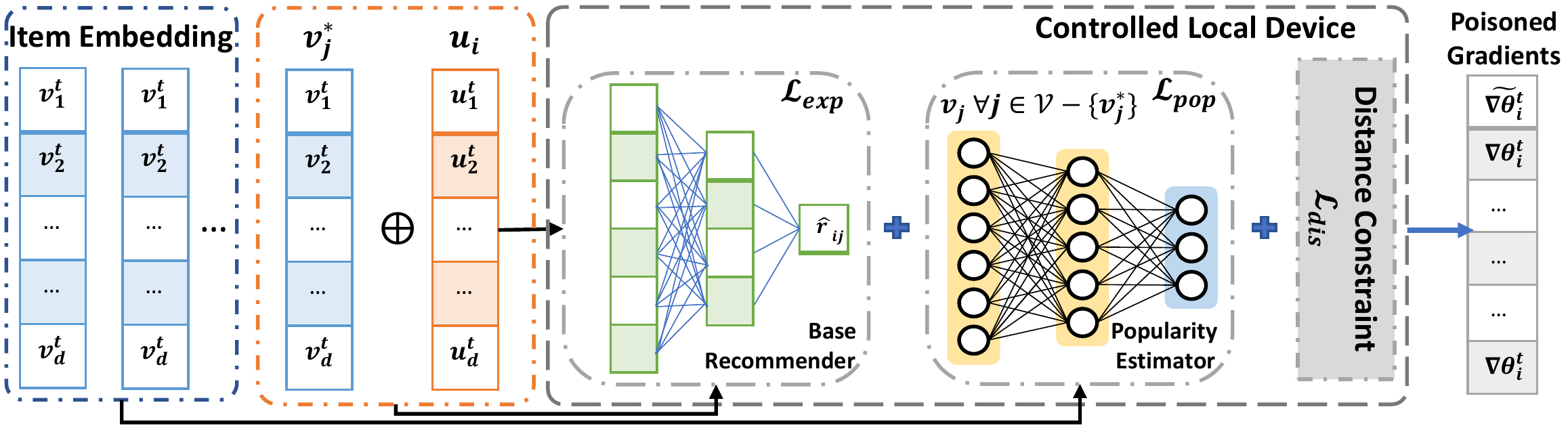}
\vspace{-2em}
\caption{Overview of PipAttack}
\label{fig:PipAttack}
\vspace{-2em}
\end{figure}
In this section, we introduce our proposed PipAttack in details.
\vspace{-0.5em}
\subsection{Base Recommender}\label{sec:base_rec}
Federated learning is compatible with the majority of latent factor models. Without loss of generality, we adopt neural collaborative filtering (NCF)~\cite{he2017neural}, a performant and widely adopted latent factor model, as the base recommender of FedRec. In short, NCF extends CF by leveraging an $L$-layer feedforward network $FFN(\cdot)$ to model the complex user-item interactions and estimate $\hat{r}_{ij}$:

\begin{equation}
    \hat{r}_{ij} = \sigma(\mathbf{h}^{\top}FFN(\mathbf{u}_i \oplus \mathbf{v}_j)),
\end{equation}
where $\textbf{u}_i, \textbf{v}_j \in \mathbb{R}^d$ are respectively user and item embeddings, $\oplus$ is the vector concatenation, $\mathbf{h} \in \mathbb{R}^{d_L}$ denotes the projection weights that corresponds to the $d_L$-dimensional output of the last network layer, and $\sigma$ is the sigmoid function that rectifies the output to fall between 0 and 1. Meanwhile, it is worth noting that our attack method is generic and applicable to many other recommenders like feature-based \cite{rendle2011fast} and graph-based \cite{he2020lightgcn} ones.
\vspace{-0.5em}
\subsection{Prior Knowledge of PipAttack}
In summary, the parameters $\Theta$ to be learned in FedRec are: the projection vector $\textbf{h}$, all weights and biases in the $L$-layer FFN, as well as embeddings $\textbf{u}_i$ and $\textbf{v}_j$ for all users and items. Meanwhile, an important note is that, as user embeddings are regarded highly sensitive as they directly reflect users' personal interests, they are commonly disallowed to be shared across users in federated recommenders \cite{qi2020privacy,chai2020secure} for privacy reasons. In FedRec, this is achieved by withholding $\textbf{u}_i$ and its gradients on every user device, and the user embeddings are updated locally. Hence, the gradients of $\textbf{u}_i$ are excluded from both $\nabla\Theta$ and $\widetilde{\nabla\Theta}$ that will be respectively submitted by benign and malicious users during the update of FedRec.

To perform poisoning attacks on FedRec, traditional attack approaches designed for centralized recommenders are inapplicable. This is due mainly to the prior knowledge that needs to be accessible to the adversary, e.g., all user-item interactions \cite{fang2018poisoning,zhang2020practical} and even other benign users' embeddings \cite{fang2020influence,lin2020attacking}, which becomes an ill-posed assumption for federated settings as most of such information is retained at the user side. In short, FedRec substantially narrows down the prior knowledge and capability of an attack model, which is restricted to the following:
\begin{itemize}[leftmargin=*]
	\item[I.] The adversary can access the global model $\Theta_t$ at any iteration $t$, excluding all benign users' embeddings, i.e., $\{\textbf{u}_i\}_{\forall u_i \in \mathcal{U}\backslash\mathcal{U}^*}$.
	\item[II.] The adversary can access and alter all malicious users' local models and their gradients.
	\item[III.] The adversary knows the whole item set (not interactions) which is commonly available on any e-commerce platform, as well as side information that reflects each item's popularity. We will expand on this in Section \ref{sec:pop_obf}.
\end{itemize}

In what follows, we present the key components of PipAttack as shown in Fig~\ref{fig:PipAttack} for learning all deceptive gradients $\{\widetilde{\nabla \Theta_i^t}\}_{\forall u_i \in \mathcal{U}^*}$, which induces FedRec to promote the target item $v^*_j$.
\vspace{-0.5em}
\subsection{Explicit Promotion}
Like many studies on poisoning attacks on recommender systems, our adversary's goal is to manipulate the learned global model such that the generated recommendation results meet the adversary's demand. Essentially, to give the target item $v^*_j$ more exposures (i.e., to be recommended to more users), we need to raise the ranking score $\hat{r}_{ij}$ whenever a user $u_i$ is paired with $v^*_j$. As a minimum requirement, we need to ensure all malicious users can receive $v^*_j$ in their top-$K$ recommendations. With the adversary's control over a group of malicious users $\mathcal{U}^*$, we can explicitly boost the ranking score of $v^*_j$ for every $u_i^* \in \mathcal{U}^*$ via the following objective function:

\begin{equation}\label{eq:L_exp}
\resizebox{0.45\linewidth}{!}{$
     \mathcal{L}_{exp} = -\sum_{u^*_i \in \mathcal{U}^*, v_j^*} \log \hat{r}_{ij},$}
\end{equation}
which encourages a large similarity score between every malicious user and the target item. Theoretically, this mimics the effect of inserting fake interaction records (i.e., data poisoning) with the target item via malicious users, which can gradually drive the CF-based global recommender to predict positive feedback on $v_j^*$ for other benign users who are similar to $u_i^* \in \mathcal{U}^*$.
\vspace{-0.5em}
\subsection{Popularity Obfuscation}\label{sec:pop_obf}
Unfortunately, Eq.(\ref{eq:L_exp}) requires the adversary to manipulate a relatively large number of malicious users in order to successfully and efficiently poison FedRec. Otherwise, after aggregating the local gradients from sampled users, the large user base of a recommender system can easily dilute the impact of deceptive gradients uploaded by a small group of $\mathcal{U}^*$. In this regard, on top of explicit promotion, we propose a more cost-effective tactic in PipAttack from the popularity perspective. As discussed in Section \ref{sec:intro}, it is well acknowledged that CF-based recommenders are intrinsically biased towards popular items that are frequently visited~\cite{abdollahpouri2019unfairness}, and FedRec is no exception. Compared with long-tail/unpopular items, popular items are more widely and frequently trained in a recommender, thus amplifying their likelihood of receiving a larger ranking score and gaining advantages in exposure rates.

Hence, we make full use of the inherent popularity bias rooted in recommender systems, where we aim to learn deceptive gradients that can trick FedRec to `mistake' $v^*_j$ as a popular item. In a latent factor model like FedRec, this can be achieved by poisoning the item embeddings in the global recommender, so that embedding $\textbf{v}^*_j$ is semantically similar to popular items in $\mathcal{V}$. Intuitively, if a popular item and $v^*_j$ are close to each other in the latent space, so will their ranking scores for the same user. Given the existence of popularity bias, $\textbf{v}^*_j$ will be promoted to more users' recommendation lists. 

\textbf{Availability of Popularity Information.} Our popularity obfuscation strategy needs prior knowledge about items' popularity. However, directly obtaining such information via user-item interaction frequencies is infeasible as it requires access to the entire dataset. Fortunately, though the user behavioral data is protected, the prosperity of online service platforms has brought a wide range of visible clues on the popularity of items. For example, hot tracks are always displayed on music applications (e.g., Spotify) without revealing identities of their listeners, and e-commerce sites (e.g., eBay) records the sales volume of each product while keeping the purchasers anonymized. Furthermore, as PipAttack is aware of the item set, item popularity can be easily looked up on a fully public platform (e.g., Yelp). Thus, popularity information can be fetched from various public sources to assist our poisoning attack, and the prerequisites of FedRec remain intact.

\textbf{Popularity Estimator.} The most straightforward way to facilitate such popularity obfuscation is to incorporate a distance metric to penalizes any $\widetilde{\nabla\Theta}_i$ that enlarges the distance between $v^*_j$ and a randomly sampled popular item $v_{j'}$. However, being a personalized model, whether an item can be recommended to $u_i$ is not only determined by its popularity, but also the user-item similarity. Therefore, such constraints will work the best only if the selected $v_{j'}$ accounts for each user's personal preference, which is infeasible due to the fact that all benign users within $\mathcal{U}\backslash\mathcal{U}^*$ are intransparent to the adversary. In PipAttack, we propose a novel popularity estimator-based method that collectively engages all the popular items in the item set. Specifically, we can assign each item a discrete label w.r.t. its popularity level obtained via public information. Then, suppose there are $C$ popularity classes, the popularity estimator $f_{est}(\cdot)$ is a deep neural network (DNN) that inputs a learned item embedding $\textbf{v}$ at iteration step $t$, and computes a $C$-dimensional probability distribution vector $\hat{\textbf{y}}$ via the final softmax layer. Each element $\hat{\textbf{y}}[c] \in \hat{\textbf{y}}$ ($c=1,2,...,C$) represents the probability that $v_i$ belongs to class $c$. We train $f_{est}(\cdot)$ with cross-entropy loss on all $(\textbf{v}, \textbf{y})$ pairs for all $v\neq v^*_j$, where $\textbf{y} = \{0, 1\}^{C}$ is the one-hot label:
\begin{equation}
\resizebox{.55\linewidth}{!}{$
    \mathcal{L}_{est} = - \sum_{\forall{v \neq v^*_j}}\sum_{c=1}^{C} \textbf{y}[c]\log \hat{\textbf{y}}[c].
    $}
\end{equation}
\textbf{Boosting Item Popularity.} Once we obtain a well-trained $f_{est}(\cdot)$, it is highly reflective of the popularity characteristics encoded in each item embedding, where items at the same popularity level will have high semantic similarity in the embedding space. So, at the $t$-th training iteration of FedRec, we aim to boost the predicted popularity of $v^*_j$ by making targeted updates on embedding $\textbf{v}^*_j\in\Theta^{t+1}$ with crafted deceptive gradients $\widetilde{\nabla\Theta}_i^{t}$. This is achieved by minimizing the negative log-likelihood of class $c_{top}$ (i.e., the highest popularity level) in the output of $f_{est}(\textbf{v}^*_j)$:
\begin{equation}\label{eq:pop_obf}
	 \mathcal{L}_{pop} = - \log f_{est}(\textbf{v}^*_j)[c_{top}], \,\, \textbf{v}^*_j \in \Theta_{t+1}.
\end{equation}
Note that $f_{est}(\cdot)$ stays fixed and will no longer be updated after the popularity obfuscation starts. Essentially, by optimizing $\mathcal{L}_{pop}$, PipAttack generates gradients $\widetilde{\nabla\Theta}_i^t$ that enforces $v^*_j$ to approximate the characteristics of all items from the top popularity group in the embedding space, resulting in a boosted exposure rate.
\vspace{-0.5em}
\subsection{Distance Constraint}
Generally, aggressively fabricated deceptive gradients $\widetilde{\nabla\Theta}_i^t$ may help the adversary achieve the attack goal with fewer training iterations, but will also incur strong discrepancies with the real gradients $\nabla\Theta_i^t$, leading to a higher chance to be detected by the central server~\cite{bhagoji2019analyzing} and harm the performance of the global model. Hence, the crafted gradients should maintain a certain level of similarity with the genuine ones. As such, we place a constraint on the distance between each $\widetilde{\nabla\Theta}_i^t$ for $u^*_j\in\mathcal{U}^*$ and all malicious users' original local gradient $\nabla\Theta_i^t$: 
\begin{equation}
\resizebox{0.72\linewidth}{!}{$
    \mathcal{L}_{dis} = \sum_{u_i \in \mathcal{U}^*} ||\widetilde{\nabla\Theta}_i^t - \frac{1}{|\mathcal{U}^*|}\sum_{u_i \in \mathcal{U}^*} \nabla\Theta_i^t||_{p},
    $}
\end{equation}
where $||\cdot||_p$ is p-norm distance and we adopt $||\cdot||_2$ in PipAttack.
\vspace{-0.5em}
\subsection{Optimizing PipAttack}
In this subsection, we define the loss function of PipAttack for model training. Instead of training each component separately, we combine their losses and use joint learning to optimize the following objective function:
\begin{equation}
    \mathcal{L} = \mathcal{L}_{exp}+\alpha \mathcal{L}_{pop}+ \gamma \mathcal{L}_{dis},
    \label{eq:final}
\end{equation}
where $\alpha$ and $\gamma$ are non-negative coefficients to scale and balance the effect of each part.
\vspace{-0.5em}
\section{Experiments}
In this section, we first outline the evaluation protocols for our PipAttacks and then conduct experiments on two real-world datasets to evaluate the performance of PipAttack. Particularly, we aim to answer the following research questions (RQs) via experiments:
\begin{itemize}
    \item[\textbf{RQ1:}] Can PipAttack perform poisoning attacks effectively on federated recommender systems?
    \item[\textbf{RQ2:}] Does PipAttack harm the recommendation performance of the federated recommender significantly?
    \item[\textbf{RQ3:}] How does PipAttack benefit from each key component?
    \item[\textbf{RQ4:}] What is the impact of hyperparameters to PipAttack?
    \item[\textbf{RQ5:}] Can PipAttack bypass defensive strategies deployed at the server side?
\end{itemize}
\vspace{-0.5em}
\subsection{Experimental Datasets}
\label{lb:data}
We adopt two popular public datasets for evaluation, namely MovieL\\
ens-1M (ML) \cite{harper2015movielens} and Amazon (AZ) \cite{he2016ups}. ML contains 1 million ratings involving 6,039 users and 3,705 movies, while AZ contains 103,593 ratings involving 13,174 users and 5,970 cellphone-related products. Following \cite{he2017neural,10.1145/3077136.3080777,10.1145/2911451.2911489}, we have binarized the user feedback, where all ratings are converted to $r_{ij}=1$, and negative instances are sampled $q=4$ times the amount of positive ones.
PipAttack needs to obtain the popularity labels of items. As described in Section \ref{sec:pop_obf}, such information can be easily obtained in real-life scenarios (e.g., page views) without sacrificing user identity in FedRec. However, as our datasets are mainly collected for pure recommendation research, there is no such side information available. Hence, we segment items' popularity into three levels by sorting all items by the amount of interactions they receive, with intervals of the top 10\% (high popularity), top 10\% to 45\% (medium popularity), and the last 55\% (low popularity). Note that we only use the full dataset once to compensate for the unavailable side information about item popularity, and the interaction records of each user are locally stored throughout the training of FedRec and poisoning attack.

\vspace{-0.5em}
\subsection{Evaluation Protocols}
Following the common setting for attacking federated models \cite{bhagoji2019analyzing}, we first train FedRec without attacks (malicious users behave normally in that period) for several epochs, then PipAttack starts poisoning FedRec by activating all malicious users. To ensure fairness in evaluation, all tested methods are asked to attack the same pre-trained FedRec model. We introduce our evaluation metrics below.
\begin{figure*}[t!]
\centering
\begin{tabular}{cccc}
\multicolumn{4}{c}{ \includegraphics[scale=0.40]{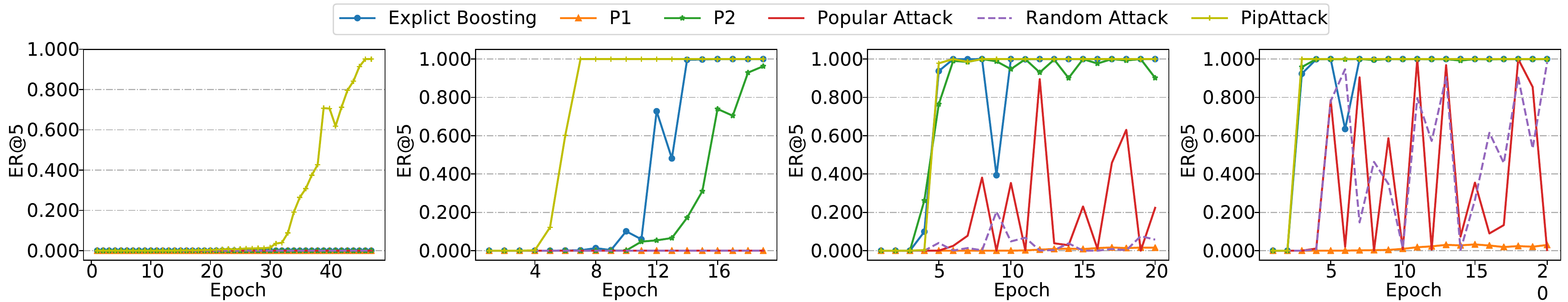}}\vspace{-0.3em}\\
\hspace{2em}\small(a) ER@5 with $\zeta = 10\%$ on ML &\hspace{0em}\small(b) ER@5 with $\zeta = 20\%$ on ML&\hspace{0.5em}\small(c) ER@5 with $\zeta = 30\%$ on ML&\hspace{0em}\small(d) ER@5 with $\zeta = 40\%$ on ML\\
\multicolumn{4}{c}{\includegraphics[scale=0.40]{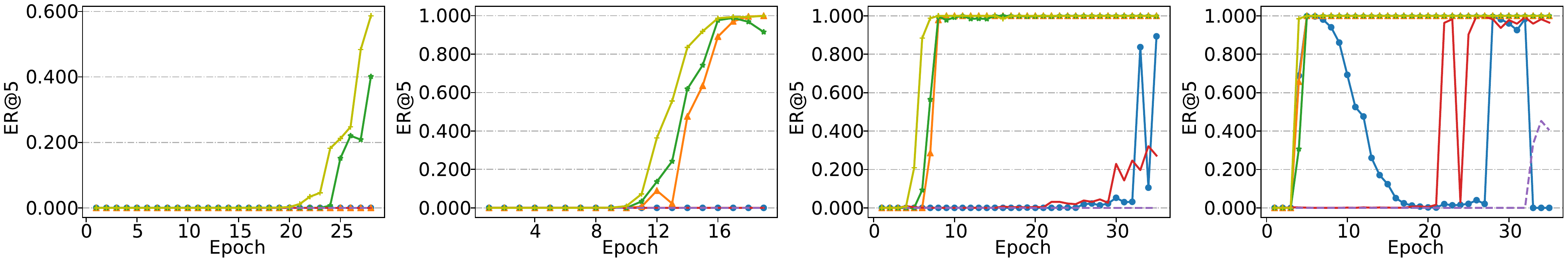}}\vspace{-0.3em}\\
\hspace{2em}\small(e) ER@5 with $\zeta = 5\%$ on AZ&\hspace{0em}\small(f) ER@5 with $\zeta = 10\%$ on AZ&\hspace{0.5em}\small(g) ER@5 with $\zeta = 20\%$ on AZ&\hspace{0em}\small(h) ER@5 with $\zeta = 30\%$ on AZ\\
\multicolumn{4}{c}{ \includegraphics[scale=0.40]{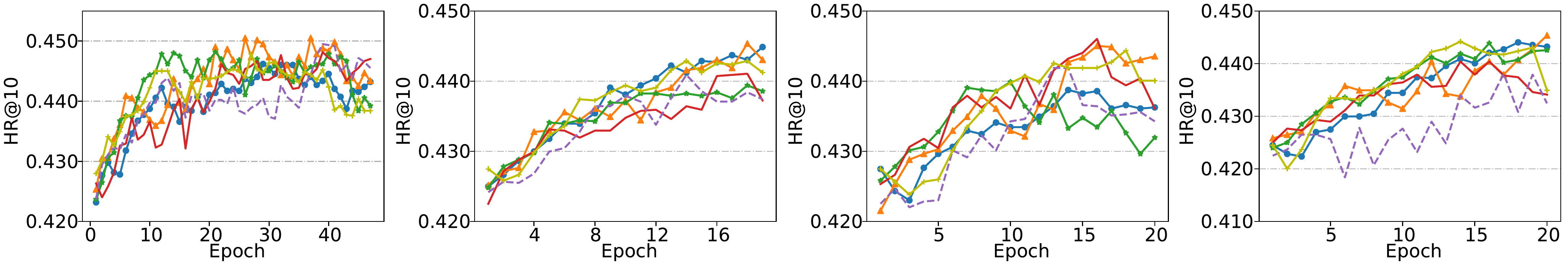}}\vspace{-0.3em}\\
\hspace{2em}\small(i) HR@10 with $\zeta = 10\%$ on ML &\hspace{0em}\small(j) HR@10 with $\zeta = 20\%$ on ML&\hspace{0.5em}\small(k) HR@10 with $\zeta = 30\%$ on ML&\hspace{0em}\small(l) HR@10 with $\zeta = 40\%$ on ML\\
\multicolumn{4}{c}{\includegraphics[scale=0.40]{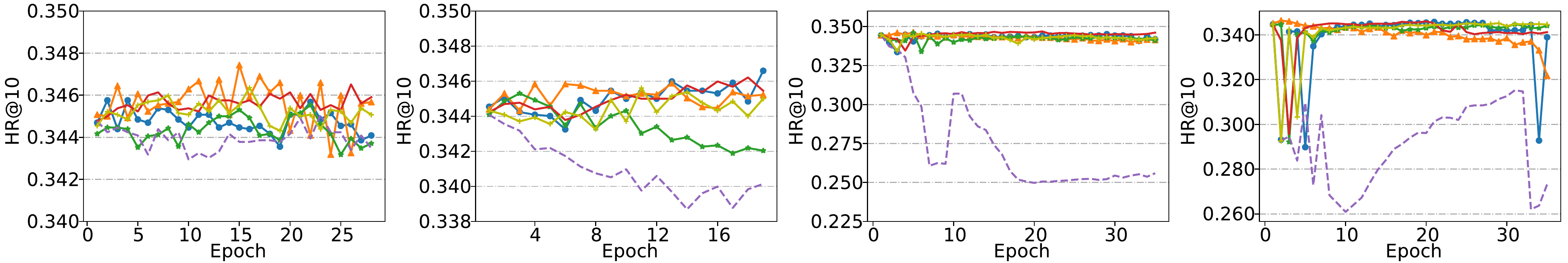}}\vspace{-0.3em}\\
\hspace{2em}\small(m) HR@10 with $\zeta = 5\%$ on AZ&\hspace{0em}\small(n) HR@10 with $\zeta = 10\%$ on AZ&\hspace{0.5em}\small(o) HR@10 with $\zeta = 20\%$ on AZ&\hspace{0em}\small(p) HR@10 with $\zeta = 30\%$ on AZ\\
\end{tabular}
\vspace{-1em}
\caption{Attack ((a)-(h)) and recommendation ((i)-(p)) results on ML and AZ. Note that the curves start from the first epoch after the attack starts.}
\vspace{-1.3em}
\label{fig:RQ1_2_result}
\end{figure*}

\textbf{Poisoning Attack Effectiveness.} We use exposure rate at rank $K$ ($ER@K$) as our evaluation metric. Suppose each user receives $K$ recommended items, then for the target item to be promoted, $ER@K$ is the fraction of users whose $K$ recommended items include the target item. Correspondingly, larger $ER@K$ represents stronger poisoning effectiveness. Notably, from both datasets, we select the least popular item as the target item in our experiments as this is the hardest possible item to promote. For each user, all items excluding positive training examples are used for ranking.

\textbf{Recommendation Effectiveness.} It is important that the poisoning attack does not harm the accuracy of FedRec. To evaluate the recommendation accuracy, we adopt the leave-one-out approach~\cite{10.1145/2911451.2911489} to hold out ground truth items for evaluation. An item is held out for each user to build a validation set. Following \cite{he2017neural}, we employ hit ratio at rank $K$ ($HR@K$) to quantify the fraction of observing the ground truth item in the top-$K$ recommendation lists. 
\vspace{-1em}
\subsection{Baselines}
We compare PipAttack with five poisoning attack methods, where the fist three are model poisoning and the last two are data poisoning methods. Notably, many recent models are only designed to attack centralized recommenders, thus requiring prior knowledge that cannot be obtained in the federated setting. Hence, we choose the following baselines that do not hold assumptions on those inaccessible knowledge:
\textbf{P1}~\cite{bhagoji2019analyzing}: This work aims to poison federated learning models by directing the model to misclassify the target input. \textbf{P2}~\cite{baruch2019little}: This is a general approach for attacking distributed models and evading defense mechanisms.
\textbf{Explicit Boosting (EB)}: The adversary directly optimizes the explicit item promotion objective $\mathcal{L}_{exp}$.
\textbf{Popular Attacks (PA)}~\cite{gunes2014shilling}: It injects fake interactions with both target and popular items via manipulated malicious users to promote the target item. 
\textbf{Random Attacks (RA)}~\cite{gunes2014shilling}. It poisons a recommender in a similar way to PA, but uses fake interactions with the target item and randomly selected items.


\subsection{Parameters Settings}
\begin{figure*}
\centering
\begin{tabular}{cccc}
\includegraphics[scale=0.65]{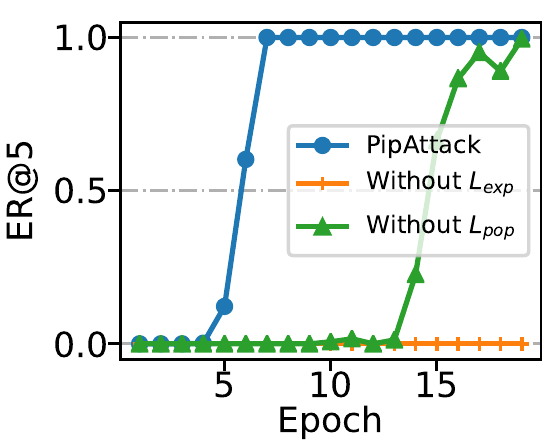}\vspace{-0.1em}&\hspace{1em}\includegraphics[scale=0.65]{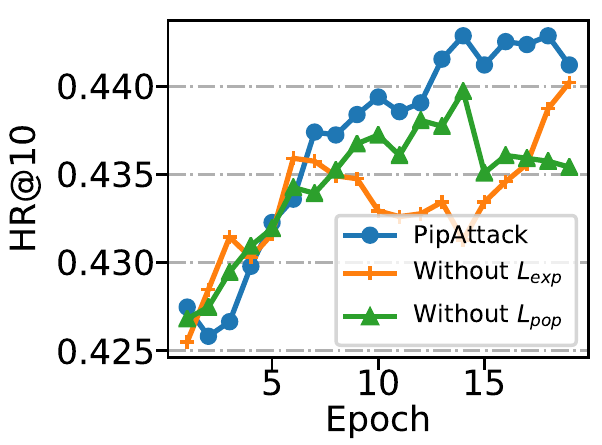}\vspace{-0.1em}&\hspace{1em}\includegraphics[scale=0.65]{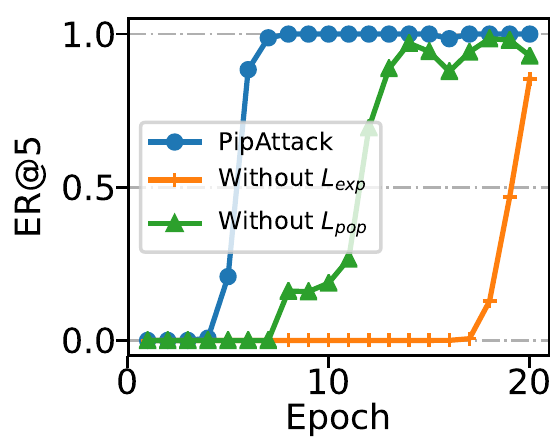}\vspace{-0.1em}&\hspace{1em}\includegraphics[scale=0.65]{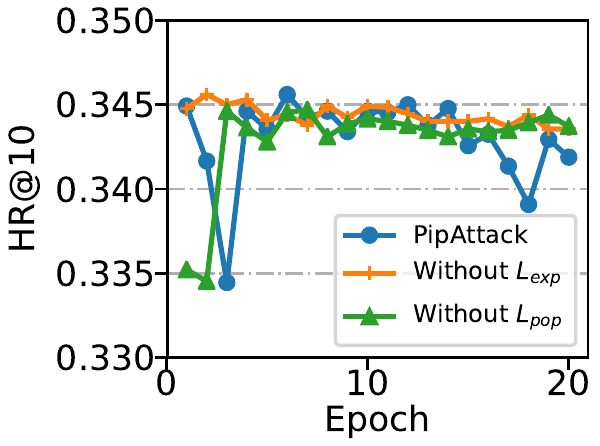}\vspace{-0.1em}\\
\hspace{2em}\small(a) ER@5 on ML&\hspace{4em}\small(b) HR@10 on ML&\hspace{4em}\small(c) ER@5 on AZ&\hspace{4em}\small(d) HR@10 on AZ\\
\end{tabular}
\vspace{-1.2em}
\caption{Ablation test with different model architectures.}
\label{fig:ablation}
\vspace{-1.2em}
\end{figure*}
\begin{figure}[t!]
\centering
\begin{tabular}{cc}
\includegraphics[width=1.2in]{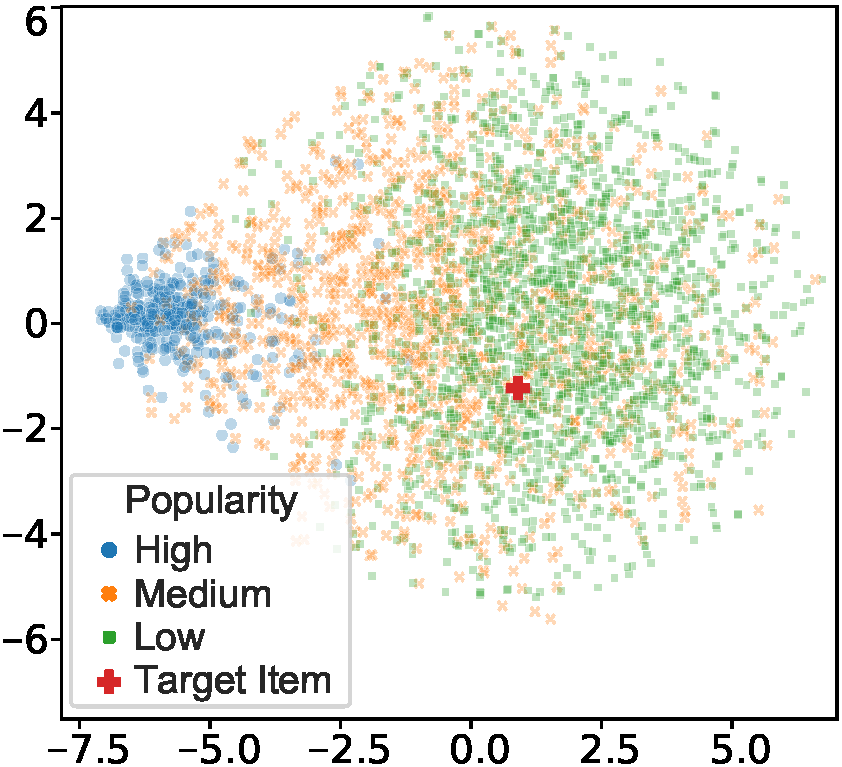}
&\includegraphics[width=1.2in]{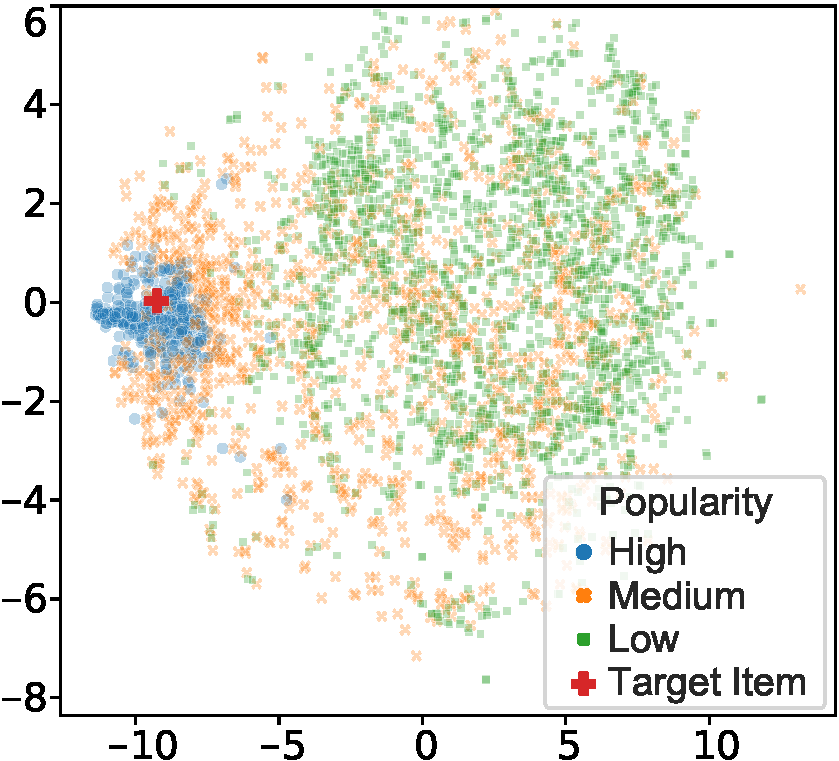}\\
\small (a) Before Poisoning Attack &\small(b) After Poisoning Attack ($ER@5=1$) \\
\end{tabular}
\vspace{-1.2em}
\caption{Visualization of item embeddings before and after being attacked by PipAttack.}
\label{fig:visual}
\vspace{-1.4em}
\end{figure}

\begin{figure}[t!]
\centering
\begin{tabular}{cc}
\hspace{-2em}\includegraphics[width=1.2in]{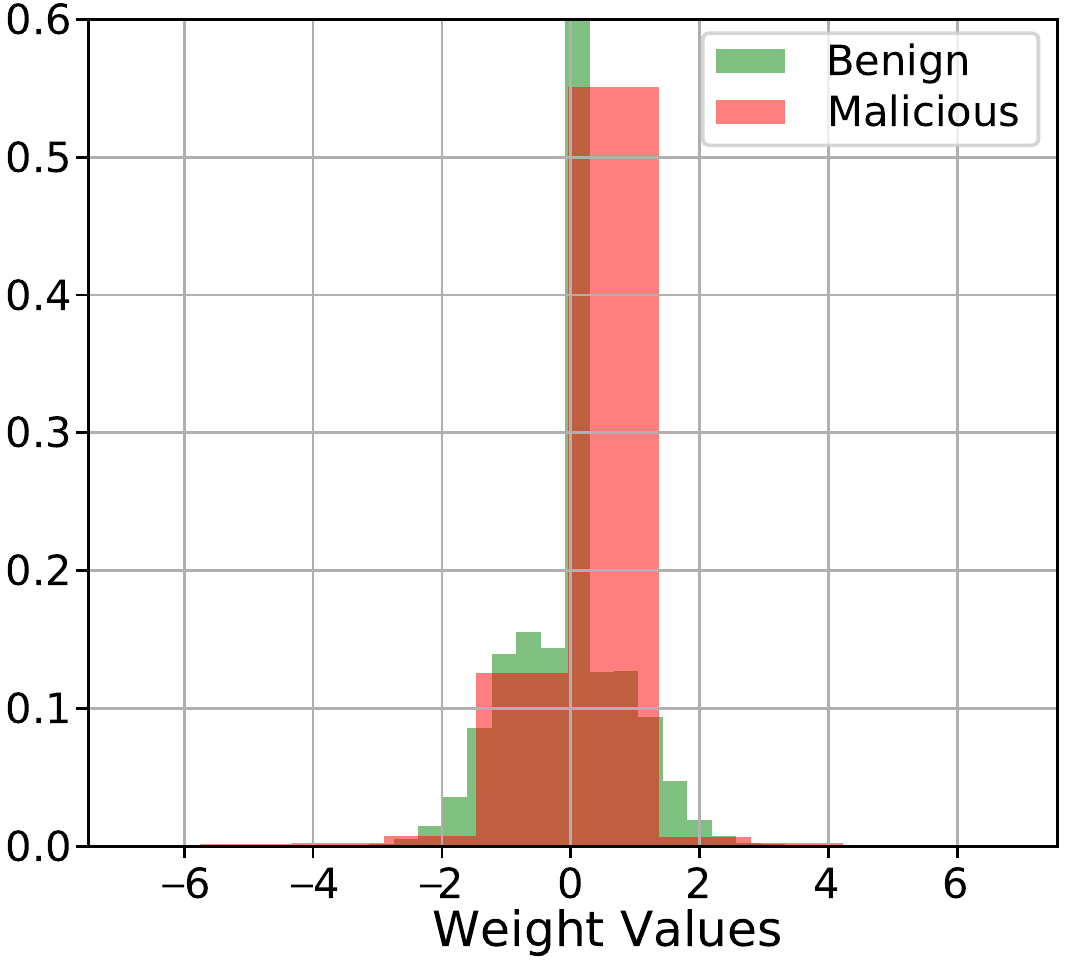}
&\hspace{-0.5em}\includegraphics[width=1.2in]{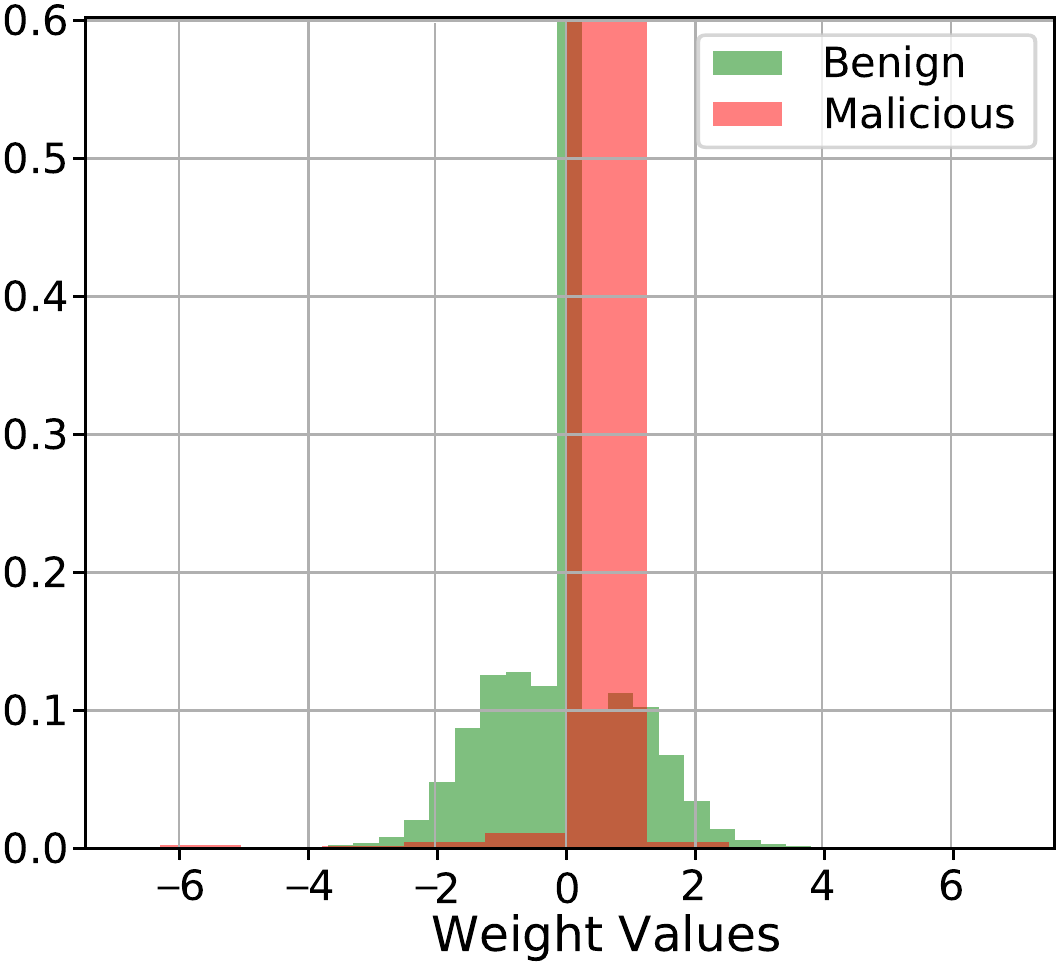}\\
\small (a) With Distance Constraint &\small (b) Without Distance Constraint.\\
\end{tabular}
\vspace{-1.4em}
\caption{Comparison of gradient distributions between benign and malicious users.}
\label{fig:distance}
\vspace{-1.4em}
\end{figure}
In FedRec, we set the latent dimension $d$, learning rate, local batch size to $64$, $0.01$ and $64$, respectively. Each user is considered as an individual device, where 10\% and 5\% of the users (including both benign and malicious users) are randomly selected on ML and AZ at every iteration. Model parameters in FedRec are randomly initialized using Gaussian distribution ($\mu = 0, \sigma = 1$). The popularity estimator $f_{est}(\cdot)$ is formulated as a 4-layer deep neural network with 32, 16, 8, and 3 hidden dimensions from bottom to top. In Pipattack, the local epoch and learning rate are respectively $30$ and $0.01$ on both datasets. We also examine the effect of different fractions $\zeta$ of malicious users with $\zeta\in\{10\%, 20\%, 30\%, 40\%\}$ on ML and $\zeta\in\{5\%, 10\%, 20\%, 30\%\}$ on AZ. For the coefficients in loss function $\mathcal{L}$, we set $\alpha=60$ when $\zeta = 10\%$ and $\alpha = 20$ when $\zeta\in\{20\%,30\%,40\%\}$ on ML, and $\alpha=10$ on AZ across all malicious user fractions. We apply $\gamma=0.0005$ on both datasets. 

\subsection{Attack Effectiveness (RQ1)}
Figure~\ref{fig:RQ1_2_result} shows the performance of $ER@5$ w.r.t. to different malicious user fractions. It is clear that PipAttack is successful at inducing the recommender to recommend the target item. In fact, the recommender becomes highly confident in its recommendation results and recommend the target item to all the users after the 40th epoch, with just a small number of malicious users (i.e., $10\%$ on ML and $5\%$ on AZ). By increasing the fraction of fraudsters, the attack performance improves significantly, as it is easier for the adversary to manipulate the parameters of the global model via more malicious users. Additionally, PipAttack outperforms all baselines consistently in terms of $ER@5$, showing strong capability to promote the target items to all users. Furthermore, as the attack proceeds, our model can consistently reach $100\%$ exposure rate in most scenarios, while there are obvious fluctuations in the performance of other baseline methods. Finally, we observe that model poisoning methods (i.e., P1, P2 and Explicit Boosting) generally perform better than data poisoning methods (i.e., Random and Popular attack), indicating the superiority of model poisoning paradigm since the global recommender can be manipulated arbitrarily for mean even with limited number of compromised users.

\subsection{Recommendation Accuracy (RQ2)}
Recommendation accuracy plays a significant role in the success of adversarial attacks. On one hand, it is an important property that the server can use to detect anomalous updates. On the other hand, only a recommender that is highly accurate will have a large user base for promoting the target item. Figure~\ref{fig:RQ1_2_result} shows the overall performance curve of FedRec after the attack starts. The first observation is that, PipAttack, which is the most powerful for item promotion, still achieves competitive recommendation results on both datasets. It proves that PipAttack can avoid harming the usefulness of FedRec. Another observation is that there is an overall upward trend for all baseline models except random attack. Finally, compared with model poisoning attacks, data poisoning attacks, especially random attack, cause a larger decrease on recommendation accuracy, which further confirms the effectiveness of model poisoning methods.

\begin{figure*}[t!]
\centering
\begin{tabular}{cccc}
\includegraphics[scale=0.60]{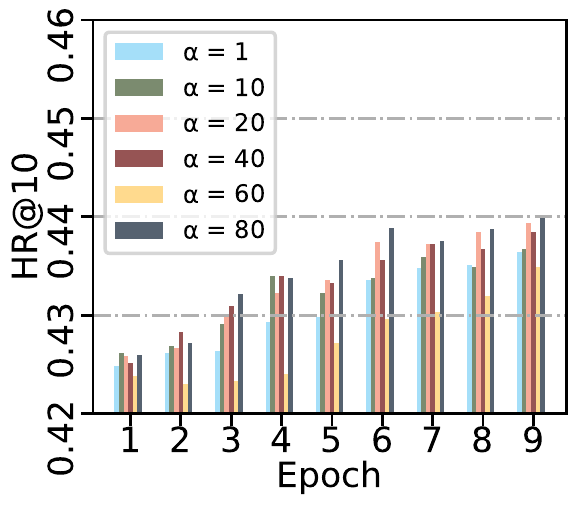}\vspace{-0.15em}&\hspace{1.7em}\includegraphics[scale=0.60]{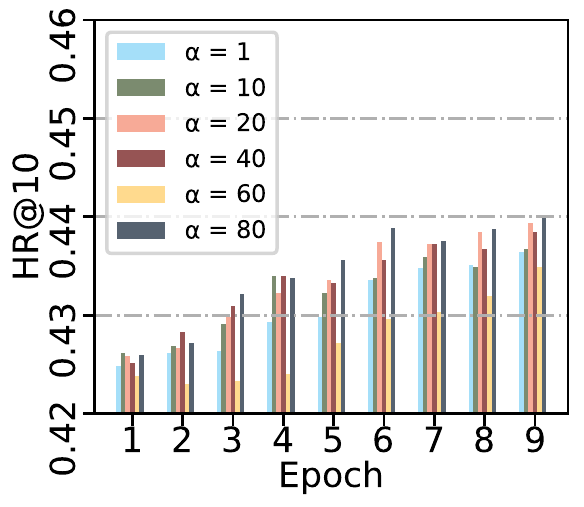}\vspace{-0.15em}&\hspace{1.7em}\includegraphics[scale=0.60]{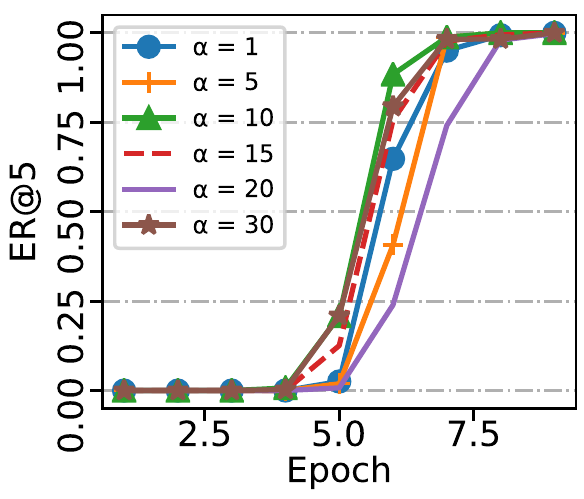}\vspace{-0.15em}&\hspace{1.7em}\includegraphics[scale=0.60]{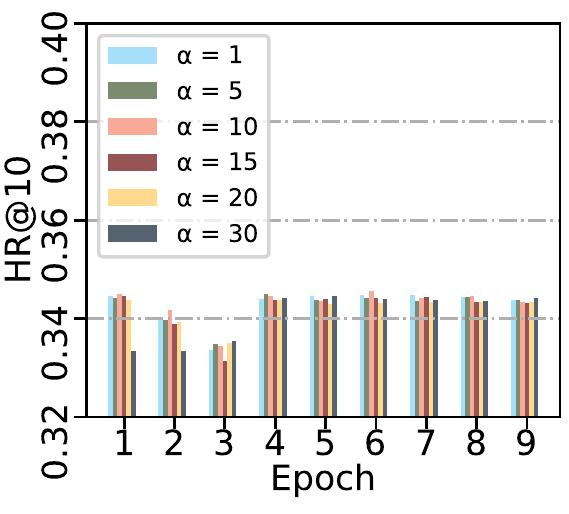}\vspace{-0.15em}\\
\hspace{0em}\small(a) ER@5 on MovieLen&\hspace{2.8em}\small(b) HR@10 on MovieLen&\hspace{3.2em}\small(c) ER@5 on Amazon&\hspace{2 em}\small(d) HR@10 on Amazon\\
\end{tabular}
\vspace{-1.2em}
\caption{Parameters sensitivity.}
\label{fig:sensitivity}
\vspace{-1.2em}
\end{figure*}

\begin{figure}[t]
\centering
\begin{tabular}{cc}
\includegraphics[scale=0.60]{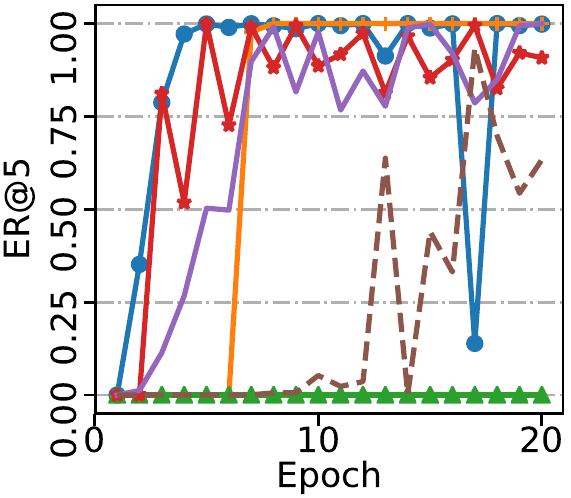}\vspace{-0.2em}&\includegraphics[scale=0.60]{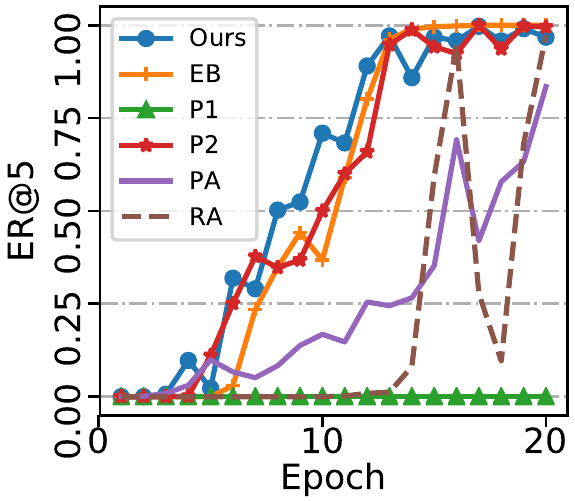}\vspace{-0.2em}\\
\small (a) With Buylan strategy.&\small(b) With Trimmed mean strategy.\\
\end{tabular}
\vspace{-1.2em}
\caption{Defense results.}
\label{fig:defense}
\vspace{-1.5em}
\end{figure}

\subsection{Effectiveness of Model Components (RQ3)}
We conduct ablation analysis to better understand the performance gain from the major components proposed in PipAttack. We set $\zeta=20\%$ throughout this set of experiments, and we discuss the key components below.

\subsubsection{Explicit Promotion Constraint.} We first remove explicit promotion constraint $\mathcal{L}_{exp}$ and plot the new results in Figure \ref{fig:ablation}. Apparently, it leads to severe performance drop on two datasets. As the main adversarial objective is to increase the ranking score for the target item, removing $\mathcal{L}_{exp}$ directly limits PipAttack's capability. In addition, the degraded PipAttack is more effective on the AZ than that on ML. A possible reason is that AZ is sparser and thus is easier to attack, despite the removal of explicit promotion constraint.

\subsubsection{Popularity Obfuscation Constraint.}
We validate our hypothesis of leveraging the popularity bias in FedRec for poisoning attack. In Figure \ref{fig:ablation}, without the popularity obfuscation constraint $\mathcal{L}_{pop}$, PipAttack suffers from the obviously inferior performance on both two datasets. For instance, $ER@5=100\%$ is respectively achieved at 17th and 14th epochs on ML and AZ, which are far behind the full version of PipAttack. It confirms that by taking advantage of the popularity bias, PipAttack can effectively accomplish the attack goal with less attack attempts. To further verify the efficacy of popularity obfuscation in PipAttack, we visualize the item embeddings via t-SNE in Figure~\ref{fig:visual}. Obviously, after the poisoning attack, the target item embedding moves from the cluster of unpopular items to the cluster of the most popular items. Hence, the use of popularity bias is proved beneficial to promoting the target item.

\subsubsection{Distance Constraint.} As introduced in ~\cite{bhagoji2019analyzing}, the server can use weight update statistics to detect anomalous updates and even refuse the corresponding updates. To verify the contribution of the distance constraint in Eq.(\ref{eq:final}), we derive one variant of PipAttack by removing the distance constraint $\mathcal{L}_{dis}$. In Figure~\ref{fig:distance}, on the ML dataset, the averaged gradients' distributions for all benign updates and malicious updates are plotted. It is clear that the deceptive gradients' distribution under the distance constraint is similar to that of benign users. In contrast, the gradients are much sparser and diverges far from benign users' gradients. Furthermore, we calculate the KL-divergence to measure the difference between two distributions in each figure. Specially, the result achieved with distance constraint ($0.005$) is much lower than the result achieved without it ($0.177$). Hence, PipAttack is stealthier, and is harder for the server to flag abnormal gradients.
\vspace{-0.5em}
\subsection{Hyperparameter Analysis (RQ4)}
\label{lab: para}
We answer RQ4 by investigating the performance fluctuations of PipAttack with varied hyperparameters on both dataset. Specifically, we examine the values of $\alpha$ in $\{1, 10, 20, 40, 60, 80\}$ on ML and $\{1, 5, 10, 15, 20, 30\}$ on AZ, respectively ($\zeta=20\%$). Note that we omit the effect of $\gamma$ as it brings much less variation to the model performance. Figure~\ref{fig:sensitivity} presents the results with varied $\alpha$. As can be inferred from the figure, all the best attack results are achieved with $\alpha = 20$ on ML and $\alpha = 10$ on AZ. Meanwhile, altering this coefficient on the attack objective has less impact on the recommendation accuracy. Overall, setting $\alpha = 20$ on ML and $\alpha = 10$ on AZ is sufficient for promoting target item, while ensuring the high-quality recommendation of FedRec.
\vspace{-0.5em}
\subsection{Handling Defenses (RQ5)}
In general, the mean aggregation rule used in FedRec assumes that the central server works in a secure environment. To answer RQ5, we investigate how our different attack models perform in the presence of attack-resistant aggregation rules, namely Bulyan~\cite{guerraoui2018hidden} and Trimmed Mean~\cite{yin2018byzantine}. Note that though Krum~\cite{blanchard2017machine} is also a widely used defensive method, it is not considered in our work due to the severe accuracy loss it causes. We benchmark all methods' performance on the ML dataset, and Figure~\ref{fig:defense} shows the $ER@5$ results with $\zeta=20\%$. The first observation we can draw from the results is that, PipAttack consistently outperforms all baselines and maintains $ER@5 = 1$, which further confirms the superiority of PipAttack. The main goal of those resilient aggregation methods is to ensure convergence of the global model under attacks, while our model's distance constraint effectively ensures this. Our experiments reveal the limited robustness of existing defenses in federated recommendation, thus emphasising the need for more advanced defensive strategies on such poisoning attacks.
\vspace{-0.5em}
\section{RELATED WORK}
A growing number of online platforms are deploying centralized recommender systems to increase user interaction and enrich shopping potential~\cite{10.1145/3442381.3449813,10.1145/3159652.3159688, 10.5555/3128489.3128560}. However, many studies show that attacking recommender systems can affect users' decisions in order to have target products recommended more often than before~\cite{lam2004shilling,christakopoulou2019adversarial, 8031037}. In~\cite{fan2021attacking}, a reinforcement learning (RL)-based model is designed to generate fake profiles by copying the benign users' profiles in the source domain, while~\cite{zhang2020practical} firstly constructed a simulator recommender based on the full user-item interaction network, and then the results of the simulator system is used to design rewards function for policy training. ~\cite{lin2020attacking, christakopoulou2019adversarial} generate fake users through Generative Adversarial Networks (GAN) to achieve adversarial intend. ~\cite{fang2018poisoning} and~\cite{fang2020influence} formulate the attack as an optimization problem by maximizing the hit ratio of the target item. However, many of these attack approaches fundamentally rely on the white-box model, in which the attacker requires the adversary to have full knowledge of the target model and dataset.

Another variant of federated recommenders are devised to inhibit the availability of dataset~\cite{qi2020privacy,wu2021fedgnn, muhammad2020fedfast,lin2020meta}. For federated recommender systems, expecting complete access to the dataset and model is not realistic.~\cite{gunes2014shilling, burke2005limited} study the influence of low-knowledge attack approaches to promote an item (e.g., random, popular attack), but the performance is unsatisfactory. Despite previous
attack methods success with various adversarial objectives under federated setting such as misclassification~\cite{bhagoji2019analyzing, baruch2019little} and increased error rate~\cite{fang2020local}, the study of promoting an item for federated recommender system still remains largely unexplored.  Therefore, we propose a novel framework that does not have full knowledge of the target model to attack under the federated setting to fill this gap.
\vspace{-0.5em}
\section{CONCLUSION}
In this paper, we present a novel model poisoning attack framework for manipulating item promotion named PipAttack. We also prove that our attack model can work on the presence of defensive protocols. With three innovatively designed attack objectives, the attack model is able to perform efficient attacks against federated recommender systems and maintain high-quality recommendation results generated by the poisoned recommender. Furthermore, the distance constraint plays an essential role in disguising malicious users as benign users to avoid detection. The experimental results based on two real-world datasets demonstrate the superiority and practicality of PipAttack over peer methods even in the presence of defensive strategies deployed at the server side.
\vspace{-0.5em}
\section{Acknowledgments}
This work is supported by Australian Research Council Future Fellowship (Grant No. FT210100624), Discovery Project (Grant No. DP190101985) and Discovery Early Career Research Award (Grant No. DE200101465).
\balance
\bibliographystyle{ACM-Reference-Format}
\bibliography{sample-base}


\end{document}